\documentclass[aps,prl,twocolumn,superscriptaddress,nofootinbib,longbibliography,10pt]{revtex4-2}

\usepackage{amsmath}
\usepackage{amsfonts}
\usepackage{enumerate}
\usepackage{amsfonts}
\usepackage[utf8]{inputenc}
\usepackage[T1]{fontenc}

\usepackage{bbm}
\usepackage{amssymb}
\usepackage{amsthm}
\usepackage{mathtools}
\usepackage{comment}
\usepackage{mathrsfs}

\usepackage[dvipsnames]{xcolor}
\definecolor{linkcolor}{rgb}{0.0,0.3,0.5}
\usepackage[unicode, colorlinks=true, linkcolor=linkcolor, citecolor=linkcolor, filecolor=linkcolor,urlcolor=linkcolor, pdfusetitle]{hyperref}

\makeatletter
\def\@fpheader{\relax}
\makeatother


\newcounter{parentsubequation}

\makeatother

\DeclareMathAlphabet{\mathbbold}{U}{bbold}{m}{n}

\begin{document}

\title{Exceptional point and hysteresis in perturbations of Kerr black holes} 

\author{João Paulo Cavalcante}
\email{joao.paulocavalcante@ufpe.br}

\affiliation{Departamento de Física, Universidade Federal de
  Pernambuco, 50670-901, Recife, Brazil}

\author{Maurício Richartz}
\email{mauricio.richartz@ufabc.edu.br}

\affiliation{Centro de Matemática, Computação e Cognição,
Universidade Federal do ABC (UFABC), 09210-580, Santo André, São Paulo, Brazil}

\author{Bruno Carneiro da Cunha}
\email{bruno.ccunha@ufpe.br}

\affiliation{Departamento de Física, Universidade Federal de
  Pernambuco, 50670-901, Recife, Brazil}

\begin{abstract}
We employ the isomonodromic method to study linear scalar massive
perturbations of Kerr black holes for generic scalar masses $M\mu$ and
generic black hole spins $a/M$. We find that the
  longest-living quasinormal mode and the first overtone coincide for
  $(M\mu)_c \simeq 0.3704981$ and $(a/M)_c\simeq 0.9994660$. We also
  show that the longest-living mode and the first overtone change
continuously into each other as we vary the parameters around the
point of degeneracy, providing evidence for the existence of a
geometric phase around an exceptional point. We interpret our findings
through a thermodynamic analogy. 
\end{abstract}

\maketitle

\noindent \textbf{\textit{Introduction --}} Black holes, according to
the uniqueness theorems of General
Relativity~\cite{Heusler_1996,Chrusciel:2012jk}, are remarkably simple
objects. In vacuum, the most general black hole is described by the
Kerr spacetime and depends only on two parameters: its mass $M$ and
its specific angular momentum
$a$~\cite{Kerr:1963ud,Wiltshire:2009zza}. Black holes are also
regarded as thermodynamic systems, with a temperature proportional to
their surface gravity and an entropy proportional to their surface
area~\cite{Bardeen:1973gs,Page:2004xp}.  

When a black hole is disturbed, it rings down through
quasinormal modes
(QNMs)~\cite{Kokkotas:1999bd,Nollert:1999ji,Berti:2009kk,Konoplya:2011qq},
characterized by complex frequencies. The real
part of $\omega$ relates to the wavelength of the oscillation, while
the inverse of the imaginary part of $\omega$ corresponds to the
characteristic decay time of the QNM. In the linear regime, each
frequency $\omega$ depends not only on the black hole itself but also
on defining parameters of the perturbation, such as the azimuthal
number $m$, the orbital number $\ell$, and the overtone number
$n$. Black hole spectroscopy uses the QNM spectrum to
infer the parameters that specify the black
hole~\cite{Dreyer:2003bv,Berti:2005ys,Berti:2007zu,Berti:2016lat,Berti:2018vdi}. This
method of describing the ringdown phase of black
hole merger events detected by gravitational wave observatories has
become particularly important in the last
decade~\cite{LIGOScientific:2020tif,Ota:2019bzl,Finch:2021qph,Cotesta:2022pci}.    

The near extremal regime $a/M \rightarrow 1$ of a Kerr black hole
provides critical insights into the nature of 
classical~\cite{Wald:1974hkz,Hubeny:1998ga,Hod:2002pm,Jacobson:2009kt,Chirco:2010rq,Natario:2016bay,Sorce:2017dst,Sasaki:1989ca,Onozawa:1995vu,Nakano:2016zvv,Hod:2008zz,Aretakis:2011gz,Lucietti:2012sf,Murata:2013daa,Gralla:2016sxp,CarneirodaCunha:2015hzd,Rosa:2009ei,Schnittman:2018ccg,Kapec:2019hro}
and quantum
gravity~\cite{Strominger:1996sh,Strominger:1997eq,Jacobson:1997ge,Guica:2008mu,Padmanabhan:2009vy,Bredberg:2009pv,Kunduri:2013gce,Matsas:2007bj,Hod:2008zza,Matsas:2009ww}. In
particular, scalar QNMs can be classified into \textit{damped modes}
(DM), for which $\mathrm{Im}(\omega)$ approaches a nonzero finite
value in the extremal limit and \textit{undamped modes}, known as
\textit{zero-damping modes} (ZDM), which are characterized by
$\mathrm{Re}(\omega) \rightarrow m/(2M)$ and $\mathrm{Im}(\omega)
\rightarrow 0$ as extremality is
approached~\cite{Andersson:1999wj,Glampedakis:2001js,Cardoso:2004hh,Hod:2011zzd,Yang:2012pj,Yang:2013uba,Cook:2014cta,Richartz:2015saa,daCunha:2021jkm}. In
fact, as the spin of a black hole approaches the extremal
limit, the imaginary part of different overtones may cross each
other~\cite{Yang:2012pj,Yang:2013uba}. Furthermore, the accumulation
of an infinite number of ZDMs at the real axis has been associated
with the horizon instability of extremal black
holes~\cite{Casals:2016mel,Richartz:2017qep,Casals:2019vdb}.  

Massive scalar fields around rotating black holes have been
investigated not only as theoretical laboratories and testbeds for
their electromagnetic and gravitational counterparts, but also in
astrophysical scenarios. They are potentially relevant to
the formation of scalar clouds, driven by superradiant
instabilities~\cite{Zouros:1979iw,Detweiler:1980uk,Dolan:2007mj,Dolan:2012yt},
around rotating black holes~\cite{Hod:2012px,Herdeiro:2014goa}, as
well as to the prospect of their use as particle
detectors~\cite{Arvanitaki:2009fg,Arvanitaki:2010sy,Brito:2015oca}. 

In this letter we systematically explore the parameter space of
massive scalar perturbations in Kerr black holes and identify 
a point of degeneracy where the fundamental QNM and its first overtone
coincide. We further identify a curve in the parameter space,
emanating from the degeneracy point, where the decay times of these
QNMs are the same. We also demonstrate that the QNMs
transform into each other when this curve is crossed in the parameter
space. This unveils the existence of hysteresis in Kerr black
holes: the result of following adiabatically a QNM in parameter space
will depend on the path taken.

We interpret our findings in terms of exceptional points of
non-hermitian systems~\cite{Kato:1995}, whose far-reaching
implications have been investigated in different dissipative systems,
specially in optics and
photonics~\cite{Berry:2004ypy,Bergholtz:2019deh,science_excep,Ding:2022juv}. Finally,
we draw an analogy of our results with the phase space of
thermodynamic systems that undergo phase transitions, where phenomena
such as critical points and level-crossing are well established.     

Throughout this work we use natural units $G=c=\hbar=1$.

\noindent \textbf{\textit{Linear perturbations --}} Linear
perturbations of a scalar field of mass $\mu$ around a Kerr black hole
are described by the Klein-Gordon equation. In Boyer-Lindquist
coordinates, the equation is separated into radial and angular
equations~\cite{Brill:1972xj}. The radial part is  
\begin{equation}
 \! \! \partial_r(\Delta\,\partial_rR) +
\left[\frac{[\omega(r^2+a^2)-
      am]^{2}}{\Delta}-\lambda-\mu^2r^2\right] \! R=0,
  \label{eq:radeq}
\end{equation}
where $\Delta = r^2-2Mr+a^2=(r-r_+)(r-r_-)$. The roots $r_+$ and $r_-$
correspond, respectively, to the event horizon and to the Cauchy
horizon of the black hole. The angular velocities $\Omega_\pm$ and the
temperatures $T_\pm$ associated with the horizons $r_\pm$ are given by 
\begin{equation}
\begin{gathered}
     \Omega_{\pm}= \frac{a}{2Mr_{\pm}}, \quad 2\pi T_{\pm} =
     \frac{r_{\pm}-r_{\mp}}{4M r_{\pm}}. 
\end{gathered}
\end{equation}        
The separation constant $\lambda\equiv \lambda_{\ell,m}$ depends on 
the angular quantum numbers $\ell$ and $m$ and its computation stems
from a Sturm-Liouville (eigenvalue) problem of the angular equation. 

In terms of $u=\cos\theta$, the angular equation is  
\begin{equation}
  \partial_u \left[(1-u^2)\partial_uS\right]+\left(\frac{m^2}{1-u^2}
    -c^2 u^2 - \Lambda \right)S=0,
  \label{eq:angeq}
\end{equation}
where $c=a\sqrt{\omega^2-\mu^2}$ and $\Lambda = \lambda + 2 a m \omega
- a^2 \omega^2$. Regularity at the poles (corresponding to $u=\pm 1$)
implies that the angular solutions are spheroidal harmonics
$S(\theta)=S_{\ell m}(\theta;c)$, with parameters $\ell$ and $m$
satisfying the constraints $\ell \in \mathbb{N}$, $m \in \mathbb{Z}$
and $-\ell \leq m \leq \ell$~\cite{10.1063/1.1705135,
  PhysRevD.73.024013}.  

The angular and radial equations can be cast as the \textit{confluent
  Heun equation} (CHE)~\cite{Batic:2007it,Fiziev:2009wn}, 
\begin{equation}
  \! \! \! \!  \frac{d^2
   y}{dz^2}+\bigg[\frac{1-\theta_0}{z}+\frac{1-\theta_{t}}{z-t}
   \bigg]\frac{dy}{dz}-\bigg[\frac{1}{4}
     +\frac{\theta_{\star}}{2z}+\frac{tc_{t}}{z(z-t)}\bigg]y=0, 
  \label{eq:heuneq}
\end{equation}
which is defined for generic complex parameters $t$, $c_{t}$, 
$\{\theta_k\}=\{ \theta_{0},\theta_{t},\theta_{\star}\}$, and a
complex independent variable $z$. For convenience, we also define
$\{\theta_k\}_-=\{\theta_0,\theta_t-1,\theta_\star+1\}$. In the theory
of ordinary differential equations on the complex plane, the CHE is
characterized by regular singular points at $z=0$ and $z=t$, where the
asymptotic behavior of the solution is polynomial, and an irregular
singular point at $z=\infty$, where the asymptotic solutions are
exponentials~\cite{ronveaux1995heun,slavianov2000special}. 

In particular, the CHE parameters for the angular equation \eqref{eq:angeq} are
\begin{equation}
\begin{gathered}
    \theta_{0}=-m, \quad \theta_{t}= m, \quad \theta_{\star}=  0, \quad  \alpha =
    \sqrt{\omega^2-\mu^2}, \\
   t = -4a\alpha, \quad  
    tc_{t}= \lambda +2(1-m)a\alpha+a^2\alpha^2,
    \label{eq:angparams}
\end{gathered}
\end{equation}
while the CHE parameters for the radial equation \eqref{eq:radeq} are
\begin{equation}
\begin{gathered}
   \theta_{0}  = \frac{-i}{2\pi T_{-}}\left(\omega-m\Omega_{-}
   \right), \quad
   \theta_{t}= \frac{i}{2\pi T_{+}}\left(\omega-m\Omega_{+} \right),
   \\
   \hspace{-0.5cm}\theta_{\star}
   =\frac{2iM(2\omega^2-\mu^2)}{\alpha}, 
   \quad t=2i\alpha (r_+-r_-), \\ 
   \hspace{-1.2cm} tc_{t}  = \lambda +
   r_+^2\mu^2-(3a^2+r_-^2+3r_+^2)\omega^2 \\ 
  \hspace{0.8cm} +i (r_--r_+-2iam+
  2i(a^2+r_+^2)\omega)\alpha \\
  \hspace{1.5cm} + i\frac{M(2\omega^2-\mu^2)}{\alpha}+
  \frac{M^2(2\omega^2-\mu^2)^2}{\alpha^2}. 
   \label{eq:parameters}
 \end{gathered}
\end{equation}

\noindent \textbf{\textit{QNMs and the isomonodromic method --}} QNMs
correspond to solutions of the radial equation \eqref{eq:radeq} that
are purely ingoing at the event horizon and purely outgoing far away
from the black hole. The calculation of QNM frequencies is typically
performed by substituting appropriate Frobenius series expansions into
the angular and radial equations and solving the resulting recurrence
relations using continued fractions. Although this technique, known as
\textit{Leaver's method}~\cite{Leaver:1985ax,Nollert:1993zz}, enjoys
fast convergence in general, it breaks down in the extremal case $a =
M$ due to the fact that the event horizon becomes an irregular
singular point of the differential equation. There are 
extensions proposed to deal with this issue, with varying degrees
of
success~\cite{Onozawa:1995vu,Richartz:2015saa,CarneirodaCunha:2015hzd,daCunha:2021jkm}.   

One of the alternatives, known as the \textit{isomonodromic
  method}~\cite{CarneirodaCunha:2015hzd,daCunha:2021jkm}, relies on
the fact that the boundary-value problems associated with the angular
and radial equations can be cast in terms of the \textit{composite
  monodromy parameters}, which in turn dictate the analytic
continuation of the solutions of the CHE as one considers paths on the
complex plane encompassing the singular points of the equation. The
main advantage of the method is the analytic control 
it provides over the dependence of the composite monodromy parameters
on the  parameters of the equation. This control ensures a smooth
limit when determining QNM frequencies in the extremal case.  

The map between the composite monodromy parameters, which we denote by
$\sigma$ and $\eta$, and the parameters of the CHE is known in mathematics
literature as the \textit{Riemann-Hilbert map} (RH map). In the
isomonodromy method this map is solved using the \textit{Painlevé V
  transcendent}~\cite{Gamayun:2013auu}. Schematically, the RH map
translates into the following pair of equations 
\begin{subequations}  \label{eq:RHmap}
\begin{gather}
  \tau_V(\{\theta_k\};\sigma,\eta;t)=0, \label{eq:RHmap:sub1}\\
  t\frac{\partial}{\partial t}\text{log} \tau_V(\{\theta_k\}_{-};\sigma-1,\eta;t)
  -\frac{\theta_0(\theta_{t}-1)}{2} = t c_{t}, \label{eq:RHmap:sub2}
  \end{gather}
  \end{subequations}
where $\tau_V$ is the \textit{tau-function} of Painlevé V, a
type of special function required to solve the class of
boundary-value problems associated with black hole QNMs. 

Given a choice of physical parameters $(a/M,M\mu,\ell,m)$, the
monodromy parameter $\sigma$  can be determined through a Floquet-type
solution of \eqref{eq:heuneq}, 
\begin{equation} \label{eq:floquet}
y(z) = e^{-\frac{1}{2}z}
z^{\frac{1}{2}(\sigma+\theta_0+\theta_{t})-1}\sum_{n=-\infty}^{\infty}
C_nz^n.
\end{equation}
Substituting the series above into \eqref{eq:heuneq}
  produces a 3-term recursion relation for $C_n$, which can be solved
  as in Leaver's method, yielding
\begin{equation}
  \sigma = \sigma(\{\theta_k\},c_t,t).
  \label{eq:aparamsigma}
\end{equation}
The parameter $\sigma$ determined this way solves
\eqref{eq:RHmap:sub2}. In fact, $\sigma$ parametrizes the complex phase of
\eqref{eq:floquet} as one continues it analytically over a path
encompassing the singular points $z=0$ and $z=t$. In contrast, the
interpretation of the monodromy parameter $\eta$ is quite
involved \cite{daCunha:2022ewy}. For our purposes, we can just define
the parameter by formally inverting the zero condition for $\tau_V$
\eqref{eq:RHmap:sub1}:
\begin{equation}
  \eta = \eta(\{\theta_k\},\sigma,t),
  \label{eq:aparameta}
\end{equation}
where $\sigma$ can be computed in terms of the CHE parameters by
\eqref{eq:aparamsigma}. For a detailed discussion, we refer
to~\cite{Cavalcante:2024kmy}.

The usefulness of the RH map stems from the fact that boundary value
problems involving the CHE can be solved by placing suitable
constraints on $\sigma$ and $\eta$. In particular, the boundary value
problem of the angular equation \eqref{eq:angeq}, \textit{i.e.}~regularity at
the poles, is solved by the requirement that the parameter $\sigma$,
as given in \eqref{eq:aparamsigma}, satisfies 
\begin{equation}
  \sigma = \theta_0+\theta_t+ 2(\ell+1)  = 2(\ell+1) ,\qquad \ell=0,1,2,\ldots,
  \label{eq:quantangular}
\end{equation}
where we have used $\theta_0$ and $\theta_t$ as defined in
\eqref{eq:angparams}, and related the parameter $\sigma$ to the angular
quantum number $\ell$. This equation relates the eigenvalues $\lambda$
and $\omega$ associated with the QNM problem. For $a\alpha \ll 1$, the
expansion of $\lambda=\lambda(\omega)$ coincides with the usual
expansion for the eigenvalue of the spheroidal
harmonics~\cite{PhysRevD.73.024013}. Similarly, the asymptotic
behavior for large $a\alpha$ can also be determined using monodromy
techniques~\cite{Casals:2018cgx,daCunha:2022ewy}. 

Concerning the radial equation, QNMs are associated with
  boundary conditions which are purely ingoing at $r=r_+$ and purely
  outgoing at $r=\infty$. As discussed in more detail in
  \cite{Cavalcante:2024kmy}, the mapping from $r$ in
  \eqref{eq:radeq} to $z$ in \eqref{eq:heuneq} depends on
  $\alpha$, and for $\mathrm{Re}\,\alpha>0$ the QNM boundary condition translates
  to
\begin{multline}
  e^{\pi i\eta}=e^{-\pi i\sigma}\frac{
    \sin\tfrac{\pi}{2}(\theta_\star+\sigma)}{
    \sin\tfrac{\pi}{2}(\theta_\star-\sigma)}\times \\
  \frac{\sin\tfrac{\pi}{2}(\theta_t+\theta_0+\sigma)
    \sin\tfrac{\pi}{2}(\theta_t-\theta_0+\sigma)}{
    \sin\tfrac{\pi}{2}(\theta_t+\theta_0-\sigma)
    \sin\tfrac{\pi}{2}(\theta_t-\theta_0-\sigma)},
  \label{eq:quantizationV}
\end{multline}
with $\{\theta_k\}$ defined in \eqref{eq:parameters}. After using
\eqref{eq:aparamsigma} and  \eqref{eq:aparameta}, the equation above
yields another constraint between the separation constant $\lambda$
and the QNM frequency $\omega$.  

\noindent \textbf{ \textit{Numerical Implementation --}} 
QNMs are characterized by the integers $\ell$ and $m$ and depend on
two dimensionless parameters, the normalized mass of the field $M\mu$
and the normalized angular momentum of the black hole $a/M$.  
We solve the coupled equations \eqref{eq:quantangular} and
\eqref{eq:quantizationV} numerically to determine the angular
eigenvalue $\lambda$ and the radial eigenvalue (i.e.~the QNM
frequency) $\omega$. For a given set of parameters, there is a
countably infinite number of frequencies $\omega$, indexed by the
non-negative integer $n$, that solve the equations.  When
$M\mu=a/M=0$, we classify these infinitely many QNMs according to the
magnitude of the imaginary part of their frequency. In other words,
$n=0$ refers to the longest living mode and $n=1$ to the first
overtone when the field is massless and the black hole is nonrotating.

In this letter, we investigate the $\ell=m=1$ QNMs across the two-dimensional parameter space $(a/M, M\mu)$, focusing on the question
of what is the longest-living mode as the mass of the
scalar field and the spin of the black hole vary. At generic points, we
follow adiabatically the roots of the RH map with
increments in $M\mu$ of order $10^{-3}$ and increments in $a/M$ of
order $10^{-5}$. Near the exceptional point (see below) and near the
extremal limit, we reduce the increments by 2 or 3 orders of
magnitude. 

We carry out the isomonodromic method using a
\href{http://www.julialang.org}{Julia} implementation of the Fredholm
determinant formulation of $\tau_V$, truncated at $N_f=64$ levels. The
determination of the parameter $\sigma$ employs a continued fraction
approximant at $N_c=128$ levels. We use the
\href{https://github.com/JeffreySarnoff/ArbNumerics.jl}{ArbNumerics}
implementation for arbitrary precision arithmetic and dense matrix
determinant calculations, working at $384$ digits. When possible, we
used both small and large $t$ expansions to ensure numerical stability
of at least $12$ digits.  Numerical codes and datasets are publicly available~\cite{codeanddata}.

\noindent \textbf{\textit{Exceptional point and ``level crossing'' --}} 
We reveal the existence of an exceptional point in the parameter
space where the fundamental QNM and its first overtone degenerate. In
perturbation theory, exceptional points arise due to non-hermiticity
of linear operators and, hence, are also known as non-hermitian
degeneracies~\cite{Kato:1995}. Even though the effective potentials
describing perturbations of Kerr black holes, such as the one
associated with the radial equation \eqref{eq:radeq}, have been
recognized as non-hermitian for a long time, to the best of our
knowledge exceptional points had not yet been identified in black
hole physics.
 
\begin{figure}[htb!]
\begin{center}
   \includegraphics[width=0.48\textwidth]{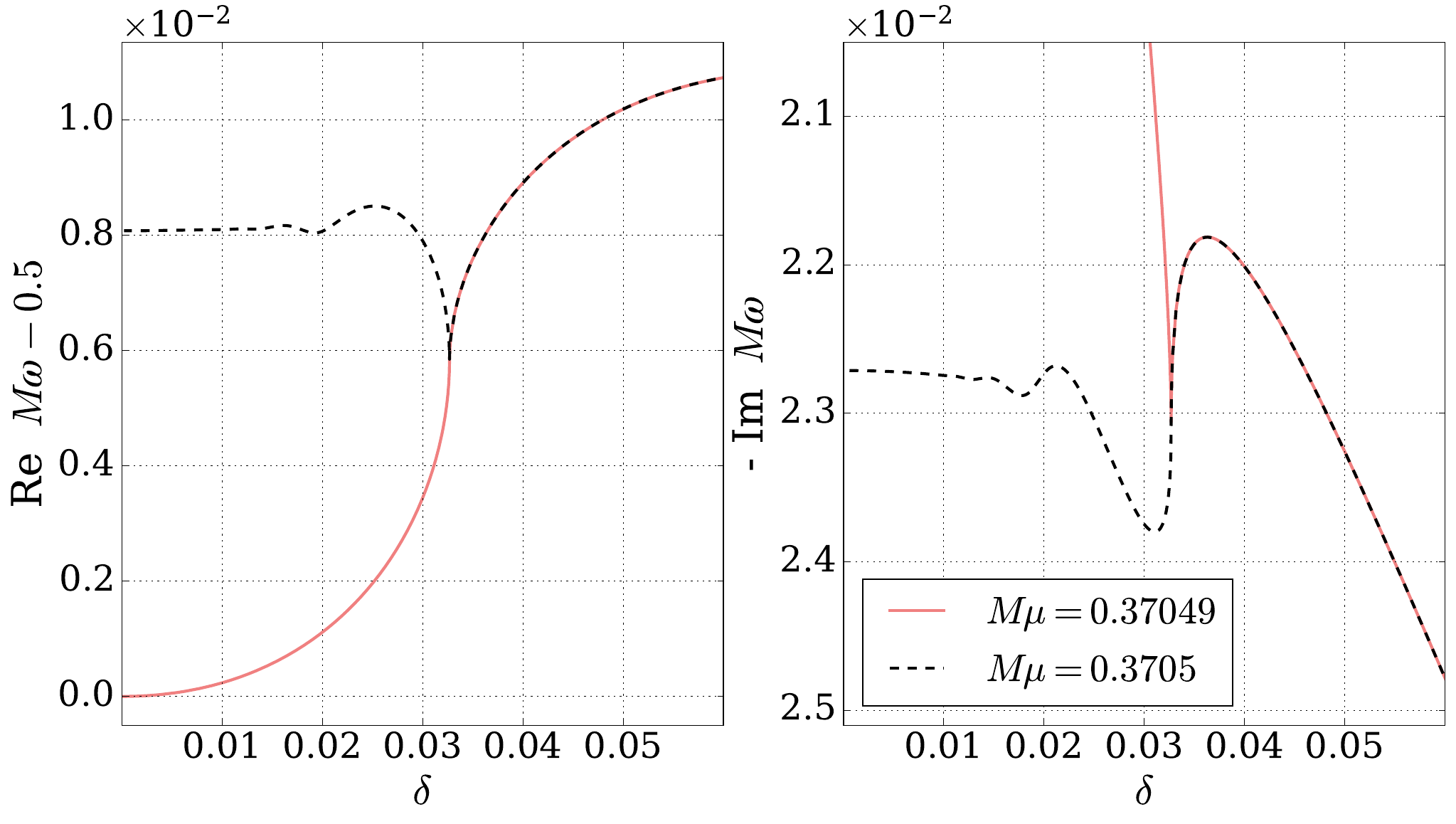}
\end{center}
\caption{The $n=0$ QNM frequency as a function of
  $\delta=\cos^{-1}(a/M)$. The solid line shows the behavior for $M\mu
  \lesssim (M\mu)_c$, whereas the dashed line corresponds to 
  $M\mu \gtrsim (M\mu)_c$. We observe the exceptional point at
  $\delta=\delta_c\simeq 0.0326823$.}   
\label{fig:s0l1trans}
\end{figure}

\begin{figure}[htb!]
  \begin{center}
    \includegraphics[width=0.48\textwidth]{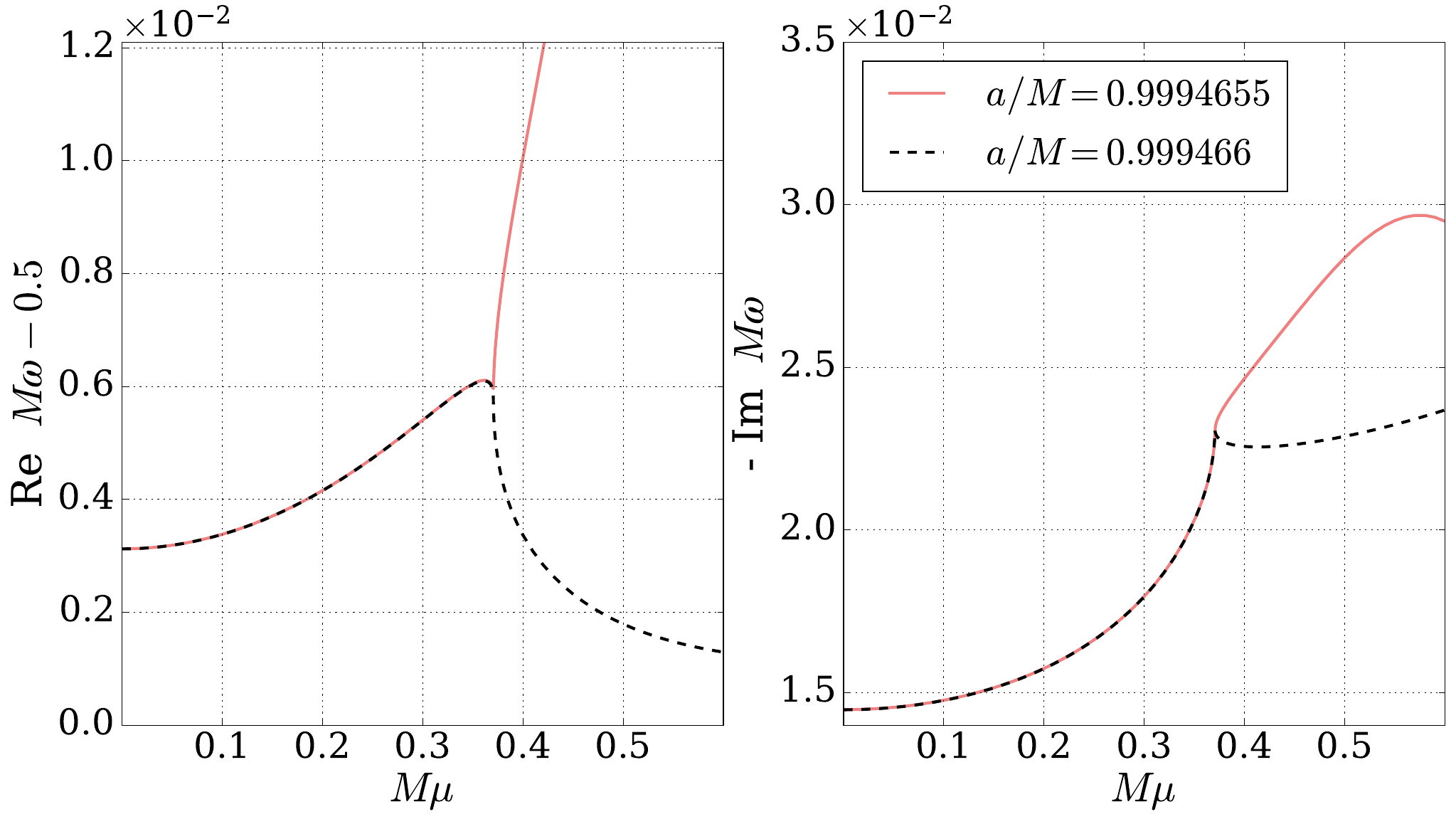}
  \end{center} 
    \caption{The $n=0$ QNM as a function of $M\mu$. The
      solid line corresponds to $a/M < (a/M)_c$, while the
      dashed curve corresponds to $a/M>(a/M)_c$. We observe the
      exceptional point at $M\mu=(M\mu)_c\simeq 0.3704981$.}  
  \label{fig:s0l1m1dampmod}
\end{figure}

By exploring the frequencies of the fundamental QNM and its first
overtone as extremality is approached at different values of the
scalar field mass, we have identified a transition from ZDMs to
DMs. Within numerical accuracy, we have found that, for
$(M\mu)_c\simeq 0.3704981$, the $n=0$ mode changes its behavior as we
increase $a/M$. For $M\mu < (M\mu)_c$, the $n=0$ mode is a ZDM, and
its frequency approaches $m/2M$ in the extremal limit. Conversely, for
$M\mu>(M\mu)_c$, the mode is a DM, meaning that the imaginary part of
its frequency does not approach zero as $a/M\rightarrow 1$. The sudden
change of behavior is shown in Fig.~\ref{fig:s0l1trans} as $a/M$
varies towards extremality. The same transition is observed in
Fig.~\ref{fig:s0l1m1dampmod}, where we vary the mass of the scalar
field for a selection of black hole spins around the critical value
$(a/M)_c$. As both figures unveil, at the critical point defined by
$M\mu=(M\mu)_c$ and $a/M=(a/M)_c\simeq 0.9994660$, the fundamental QNM 
develops a cusp. We remark that a similar behavior was observed for
charged massless perturbations of Reissner-Nordstr\"om black
holes~\cite{Cavalcante:2021scq}. 


\begin{figure}[htb!]
  \begin{center}
  \includegraphics[width=0.48\textwidth]{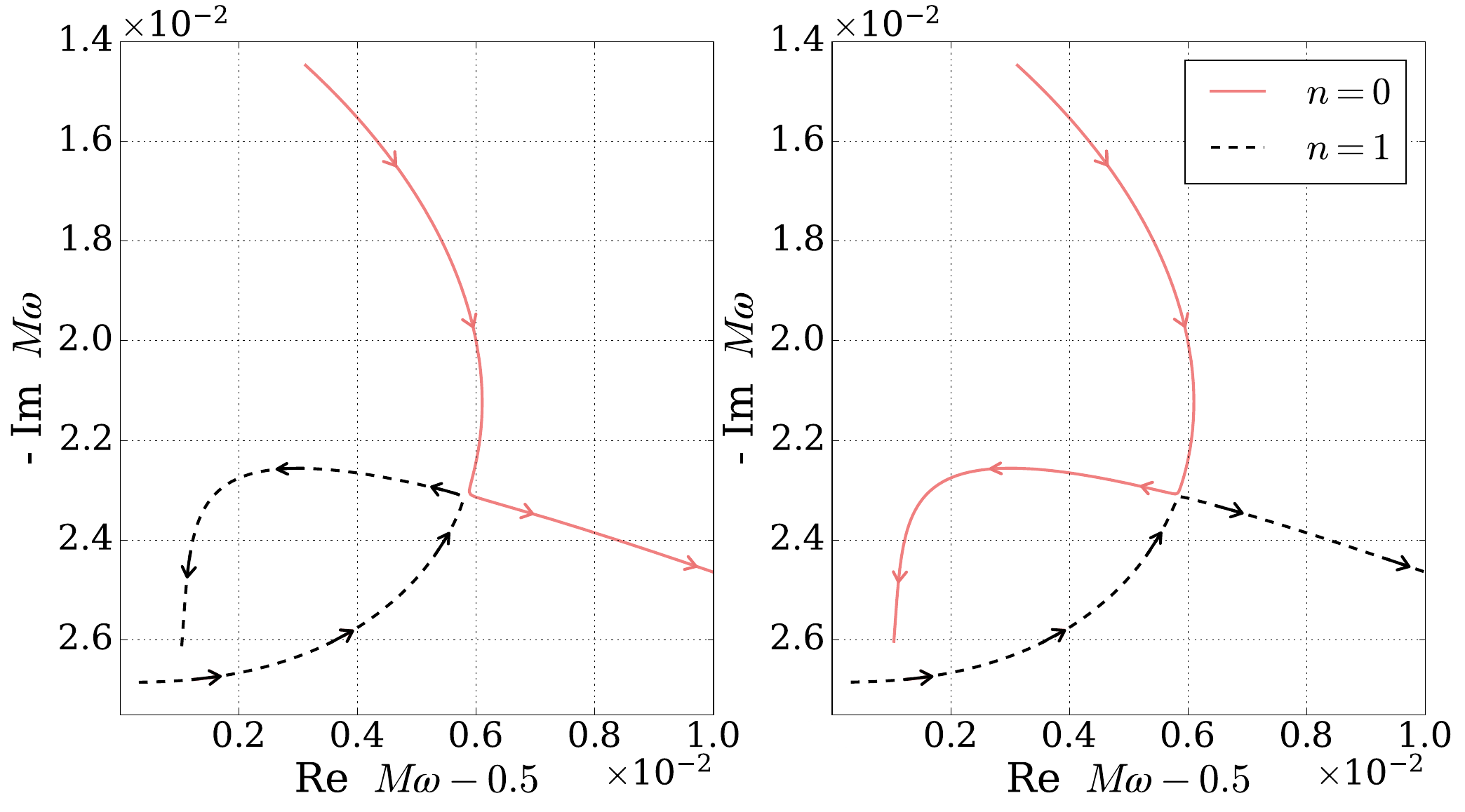}
  \end{center}
  \caption{ The $n=0$ and $n=1$ QNMs as a function of
      $M\mu$ for $a/M \lesssim (a/M)_c$ (left) and $a/M \gtrsim
      (a/M)_c$ (right). The arrows denote the direction of increasing
      $M\mu \in [0.0, 0.8]$. At $\{a/M,M\mu\}=\{(a/M)_c,(M\mu)_c\}$, the 
    QNMs become degenerate and transform into each other, resembling
    the level crossing behavior of thermodynamical systems.}
  \label{fig:s01change1}
\end{figure}

\begin{figure}[htb!]
\begin{center}
   \includegraphics[width=0.38\textwidth]{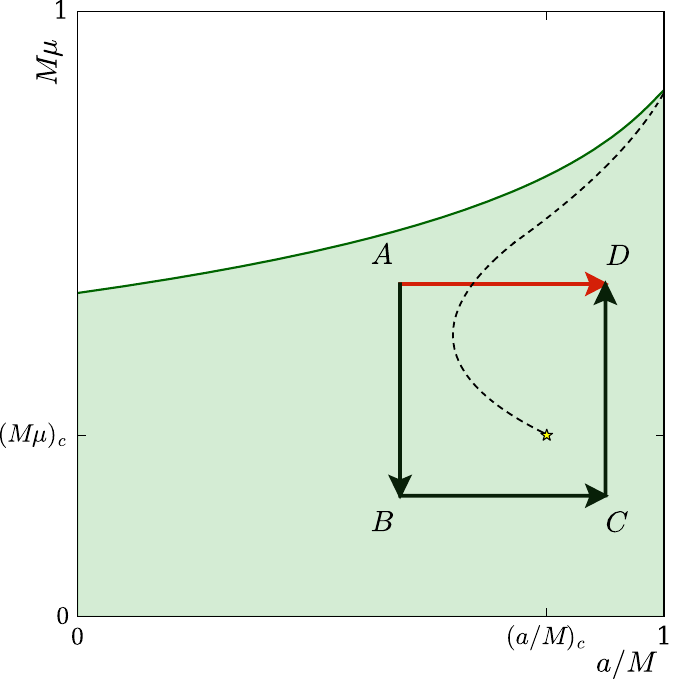}
\end{center}
\caption{Schematic representation of the parameter space associated
  with $\ell=m=1$ massive scalar perturbations in Kerr black
  holes. The star corresponds to the exceptional point where the $n=0$
  and the $n=1$ QNMs are degenerate. The dashed line correponds to the
  coexistence curve where the decay rates of these QNMs are
  equal. Deformations of a QNM along the paths $AD$ and $ABCD$ yield
  different results, characterizing the ocurrence of hysteresis. The
  curve separating the shaded (green) and white regions is characterized
  by $\mathrm{Re}\,\alpha=0$ for the $n=0$ modes.}
 \label{fig:hysteresis}
\end{figure}

The fact that the $n=0$ modes become DMs for $M\mu>(M\mu)_c$ when the 
black hole is spinning sufficiently fast implies that they may not be
the longest-living modes at every point of the parameter space. In
fact, by following the $n=0$ and the $n=1$ QNMs throughout the
parameter space, we have observed that their frequencies coincide at
the critical point, which is thus understood as an exceptional point
of a non-hermitian system.
One of the exotic features of the
exceptional point is shown in Fig.~\ref{fig:s01change1}, where we plot
both QNMs as a function of $M\mu$ for $a/M \lesssim (a/M)_c$ and $a/M \gtrsim (a/M)_c$. As the spin increases through the critical value,
we observe that the QNMs degenerate and transform into each other. 
The parameter space $(a/M,M\mu)$ is shown in
Fig.~\ref{fig:hysteresis}, where the exceptional point is represented
as a star. We have also identified a ``curve of coexistence'' emerging
from the critical point, represented as a dashed curve, where the
imaginary parts of the frequencies for the $n=0$ and the $n=1$ modes
coincide. In the accompanying paper \cite{Cavalcante:2024kmy} we
characterize the coexistence curve with greater detail. 

Note the curve separating the shaded (green) and white regions, defined by $\mathrm{Re} \, \alpha = 0$ for the $n=0$ modes, above which the QNM boundary condition no longer follows Eq.~\eqref{eq:quantizationV}. Extending the QNM analysis into the
white region is beyond the scope of this letter, but some insights are
discussed in \cite{Cavalcante:2024kmy}. When the QNMs are 
adiabatically deformed along a path that intersects the coexistence
curve, such as the path $AD$ in Fig.~\ref{fig:hysteresis}, the
longest-living mode switches to the second longest-living mode, and
vice-versa.

\noindent \textbf{\textit{Geometric phase and hysteresis --}}
So far we used $n=0$ and $n=1$ to refer to the spectrum of QNMs as
ordered by their imaginary parts in the massless Schwarzschild
configuration. The presence of an exceptional point means that following adiabatically either mode may yield different results depending on the path taken in parameter space. The path dependence establishes the existence of non-contractible curves in parameter space and ensures the existence of the
exceptional point, whose numerical evidence we discussed above.

To illustrate this, we have adiabatically followed some
of the modes around a closed path in parameter space containing the
exceptional point. We have found that, even if the starting and ending
points of the parameter space trajectory coincide, the final QNM
frequency differs from the corresponding initial value, signalling the
existence of a non-trivial geometric phase. Explicitly, consider the
rectangle $ABCDA$ shown in Fig.~\ref{fig:hysteresis}, whose vertices
$(a/M,M\mu)$ are $A=(0.999,0.45)$, $B=(0.999,0.3)$, $C=(0.9999,0.3)$
and $D=(0.9999,0.45)$. We start with the longest-living mode at $A$, whose
frequency is $\omega_{\alpha}(A)=0.514974-0.026418i$. Changing the parameters
continuously along the path $ABCD$, and thus avoiding the
  coexistence curve, yields the longest-living mode at $D$,
with frequency $\omega_{\alpha}(D) = 0.500386 - 0.009456i$. On the other hand,
deforming the QNM through the straight line $AD$, 
 and thus crossing the coexistence curve, results in the second
longest-living mode at $D$, with frequency $\omega_{\beta}(D) = 0.514827 -
0.026255i$. If we repeat the procedure starting with the second
longest-living mode at $A$, whose frequency is $\omega_{\beta}(A)=0.503306 -
0.032998i$, the results are reversed: the straight line leads to the
longest-living mode at $D$, with frequency $\omega_{\alpha}(D)$, whereas the
path that avoids the coexistence curve leads to the second
longest-living mode at $D$, with frequency $\omega_{\beta}(D)$. To corroborate
our findings, we also implemented Leaver's method~\cite{codeanddata}, as outlined in
\cite{Konoplya:2006br,Dolan:2007mj,Konoplya:2013rxa,Siqueira:2022tbc,Richartz:2024efi},
and obtained the same numerical results. 


\noindent \textbf{\textit{Discussion --}} We have gathered strong
evidence for the existence of exceptional points and geometric phases
associated with linear massive scalar perturbations of near extremal
Kerr black holes. To support these conclusions, we have examined the
dependence of the QNM frequency on the spin of the black hole and on
the mass of the scalar field, in a way that mimics thermodynamic
systems. We have presented a short discussion of the
isomonodromic method, focusing on the existence of a degeneracy and
the possibility of hysteresis for the $\ell=m=1$ modes. The complete
technical details and results for general quantum numbers are provided
in the companion paper~\cite{Cavalcante:2024kmy}, where we discuss QNM
frequencies beyond their behaviour around the exceptional point. 

We conclude this letter by interpreting our results through the lens
of statistical mechanics. The Fredholm determinant formulation of the
Painlevé V 
tau-function \cite{Lisovyy:2018mnj} expresses
$\tau_V$ as
\begin{equation}
  \tau_V = \Upsilon\det(\mathbbold{1}-\mathsf{K}(\omega)),
  \label{eq:fredholm}
\end{equation}
where $\mathsf{K}$ is an integral operator and $\Upsilon$ is a
function of $t$ that vanishes only at the critical points $t=0$ and
$t=\infty$ (corresponding to $\omega=0$ and $\omega=\infty$,
respectively). The specific forms of $\Upsilon$ and $\mathsf{K}$ are
not essential for this discussion and can be found in 
\cite{daCunha:2021jkm}. The key aspect is that identifying non-trivial
zeros of $\tau_V$, which are associated with the calculation of the
QNM frequencies through
\eqref{eq:aparamsigma}-\eqref{eq:quantizationV}, reduces to computing
zeros of the determinant in \eqref{eq:fredholm}.    

At this point, there is a clear analogy between solving for the zeros
of the determinant in \eqref{eq:fredholm} and solving the secular
equation to find the eigenvalues of the transfer matrix in a
statistical mechanical system. Futhermore, describing the ringdown of
the black hole through the longest-living frequency of
$\mathsf{K}(\omega)$ can be understood as an analogous procedure to
keeping the largest eigenvalue of the transfer matrix. In statistical
mechanics, the latter procedure amounts to computing the
free energy of the system in the thermodynamical limit. 

We posit that the existence of the exceptional point and hysteresis
phenomena we reported for Kerr black holes is analogous to the
behavior of a thermodynamical system near a critical point. The
critical point is the end of a curve of coexistence between phases,
where there is a discontinuity of the free energy of the
system. Likewise, in our coexistence curve the imaginary parts of the
frequency match but the real part does not. In the thermodynamical
system, continuing the curve beyond the critical point reveals a
cross-over where the phases trasmute smoothly into one another,
without any apparent discontinuity. We observe the same behavior by
taking the $ABCD$ path in Fig.~\ref{fig:hysteresis}, as if
$\mathsf{K}(\omega)$ were a non-linear version of the transfer matrix
of a statistical mechanical system.

Level-crossing and cross-overs are generic features of phase
transitions in thermodynamics, as demonstrated in standard textbook
treatments of the van der Waals fluid~\cite{Callen:1991}. We have
presented compelling evidence, through the specific example of massive
scalar fields in Kerr black holes, that gravitational systems can
display the same behavior. Although localized in physical configuration space, these phenomena undoubtedly impact the ringdown phase and the overall behavior of massive fields when the relevant conditions are met. These topics present exciting opportunities for future research.   

It is remarkable that the critical mass aligns with the
regime where superradiant instabilities are most
efficient, \textit{i.e.}~$(M\mu)_c  \sim  \mathcal{O}(1)$~\cite{Brito:2015oca}. In particular, putative bosonic
condensates growing from these instabilities can emit
gravitational waves through several channels (\textit{e.g.}~continuous emission arising from annihilation of the boson field or
level transitions between states with different overtone numbers). Such
processes (superradiant instabilities and gravitational wave emission)
change the mass and the spin of the black hole, drawing
paths in the parameter space (as in Fig.~\ref{fig:hysteresis}) that
may cross a coexistence curve and encompass an exceptional point. 

\noindent \textbf{Note added.} After completing this work, we learned
of a recent preprint by Motohashi~\cite{Motohashi:2024fwt}, which
interprets an avoided
crossing~\cite{Dias:2021yju,Dias:2022oqm,Davey:2023fin} near a
degeneracy point as the mechanism underlying resonances between
gravitational QNMs. 

\noindent \textbf{\textit{Acknowledgements --}} The authors thank
André Lima for comments and suggestions on the
manuscript. M.~R.~acknowledges partial support from the Conselho
Nacional de Desenvolvimento Cient\'{i}fico e Tecnol\'{o}gico (CNPq,
Brazil), Grant 315991/2023-2, and from the S\~ao Paulo Research
Foundation (FAPESP, Brazil), Grant 2022/08335-0. 

\bibliography{prl_resub}

\end{document}